\documentclass[submission, PhysLectNotes]{SciPost}
\hypersetup{
    colorlinks,
    linkcolor={red!50!black},
    citecolor={blue!50!black},
    urlcolor={blue!80!black}
}

\usepackage[bitstream-charter]{mathdesign}
\usepackage{float}
\DeclareSymbolFont{usualmathcal}{OMS}{cmsy}{m}{n}
\DeclareSymbolFontAlphabet{\mathcal}{usualmathcal}

\newcommand{\la}{\langle}
\newcommand{\ra}{\rangle}
\newcommand{\mc}{\mathcal}

\def\dbar{{\mathchar'26\mkern-14.5mu d}}

\begin{document}

\begin{center}{\Large \textbf{A brief tutorial on information theory}}\end{center}

\begin{center}
Tarek Tohme$^a$ and William Bialek$^b$ 
\end{center}
 
 \begin{center}
 $^a$Laboratoire de Physique de l’\'Ecole Normale Supérieure,  CNRS,\\ PSL University,
Sorbonne Universit\'e, and Universit\'e de Paris, 75005 Paris, France \\
 {\small \sf tarek.tohme@phys.ens.fr}
\\
$^b$Joseph Henry Laboratories of Physics and Lewis--Sigler Institute for Integrative Genomics,   Princeton University, Princeton, NJ 08544 USA
\\
 {\small \sf wbialek@princeton.edu}
 \\
\end{center}

\begin{center}
\today
\end{center}

\section*{Abstract}
{\bf
At the 2023 Les Houches Summer School on Theoretical Biological Physics, several students asked for some background on information theory, and so we added a tutorial to the scheduled lectures.  This is largely a transcript of that tutorial, lightly edited. It covers basic definitions and context rather than detailed calculations.  We hope to have maintained the informality of the presentation, including exchanges with the students, while still being useful.
}

\vspace{10pt}
\noindent\rule{\textwidth}{1pt}
\tableofcontents\thispagestyle{fancy}
\noindent\rule{\textwidth}{1pt}
\vspace{10pt}

\section{Prologue}

Evidently a short tutorial is not a substitute for a full course on information theory.   It still is very useful to read Shannon's original paper \cite{shannon_48},  and the  standard textbook is by Cover and Thomas \cite{cover+thomas_91}.  A version aimed more at the physics community is by M\'ezard and Montanari \cite{mezard+montanari_09}, and there is  a fuller account of these ideas in the context of biophysics \cite{bialek_12}.  All of these texts provide much more than what you need to make sense out of the main lectures \cite{bialek_24}, but perhaps you will want to explore more deeply.

This tutorial was given by WB, and transcribed by TT.  We have smoothed things out in a few places, but tried to keep the flavor of the original, in particular not shifting from ``I'' to the editorial ``we.''  Special thanks to the students who spoke up with questions.

\section{Some basics of information theory, with dialogue}

%Almost all statistical mechanics textbooks explain that the entropy of a gas measures our lack of information about the microscopic state of the mol\-e\-cules, but often this connection is left a bit vague or qualitative. Shannon proved a theorem that makes the connection precise: entropy is the unique measure of available information consistent with certain simple and plausible requirements. Further, entropy also answers the practical question of how much space we need to use in writing down a description of the signals or states that we observe.  

In most statistical mechanics textbooks, the entropy of a gas is interpreted as a measure of our ignorance, or lack of  information about the microscopic state of the molecules, but the reader may be left wondering if this is  more than a metaphor. Shannon proposed in 1948 a few simple and reasonable constraints  that any general measure of information should obey, and he then proved that entropy is the unique function that satisfies these constraints. Further, Shannon's entropy also solved the practical problem of finding the least amount of space we need to write down a description of states or signals we observe.

If we hear the answer to some question, we feel that we have ``gained some information.'' But how much, precisely, have we gained?  For starters, we could list  the possible answers to that question, and number them ${n} = 1,\, 2,\, \cdots ,\, N$. We also know that not all possible answers are equally likely to be the correct answer, and we can express this knowledge as a probability $p_{n}$ for each answer $n$ to be correct.

A general measure of information can't depend on the details of the question and answers.  Shannon proposed that the average information gain should depend only on the probabilities $p_{n}$ of each answer being correct, so that the information is   a function $I(\{p_{n}\})$. The difficulty is then to determine this function.\footnote{Note that Shannon's starting assumption---that only the probability distribution over the answers to our question matters in measuring the information gained---means that enumerating the possible answers must be done carefully. It also means that we cannot quantify information we would gain upon hearing an answer to our question that we could not conceive of at all. Whether this is an actual restriction is interesting to think about.}

Shannon's strategy was to impose simple, plausible constraints:
\begin{itemize}
\item In the restricted case where all $N$ answers are equally likely, then $I(\{p_{n}\}) = I(N)$ should increase monotonically with $N$---when there are more possible answers to a question, we learn more from hearing the answer.
\item If we can ask a question in two parts that are completely independent of one another, then we should be able to add the information gained from each part to find the
 information gained by answering the original question.
\item More generally, a question could be broken down into a number of conditional sub-questions: each successive question depends on the previous answer, and the probabilities of the answers are refined accordingly. In such cases the total information obtained from an answer should be the weighed sum over the entire branching tree of the information gained at each branching point.
\end{itemize}

Remarkably, these three postulates are sufficient to determine the function $I(\{p_{ n}\})$ uniquely, and it is the entropy of the probability distribution,
\begin{equation}
I(\{p_{ n}\}) = S = -k\sum_{ n = 1}^N p_{ n} \log p_{ n} .
\label{I1}
\end{equation}
You should  go through the proof yourself \cite{bialek_12,shannon_48}.  The mathematics that Shannon used was shockingly elementary.  
Equation (\ref{I1}) is not quite unique, since there is an arbitrary constant $k$ and we can also choose the base of the log; changing the base can always be absorbed into a redefinition of $k$.  This arises because nobody tells us what units to use in measuring information.  Thus if $I(\{ p_{n } \})$ is consistent with the constraints, so is the function $2I(\{p_{n}\})$.  This problem also exists when we use entropy in statistical mechanics.  One common choice among physicists is to take $k = k_B$, Boltzmann's constant, and choose natural logs.  But we also sometimes keep the combination $k_B T$, measuring temperature in energy units, and then in the definition of entropy we take $k=1$.  Chemists, in contrast, like to count moles rather than molecules, and thus define entropy with $k=R$, the gas constant; $R = N_A k_B$, with $N_A = 6.02\times 10^{23}$.

In information theory, it's also useful to keep things dimensionless, so we can always choose $k = 1$. Another question is: what base do you choose to your logarithm? A convention, as I think most of you know, is to use $\log_2$, so that the units of $I$ are bits. So what does this mean? If we agree to measure information as 
\begin{equation}
I \equiv  -\sum_{n=1}^N p_{n} \log p_{n},
\end{equation}
then with $N=2$ and $p_1 = p_2 = 1/2$, we get $I=1$ bit. So we started by saying that this information quantity will be quantifying how much information we gain when we hear the answer to a question. So the units you use are the information that you can gain by hearing the answer to a yes/no question, or any question with only two alternatives -- sort of the elementary question -- and if yes and no were equally likely answers. Why? Well, if you have two alternatives, $p_1 = p$, then $p_2 = 1-p$, if you plot the information as a function of $p$, it runs between 0 and 1, the maximum is at $1/2$ where it's one bit. \\
\begin{figure}[H]
\centering
\includegraphics[width=3in]{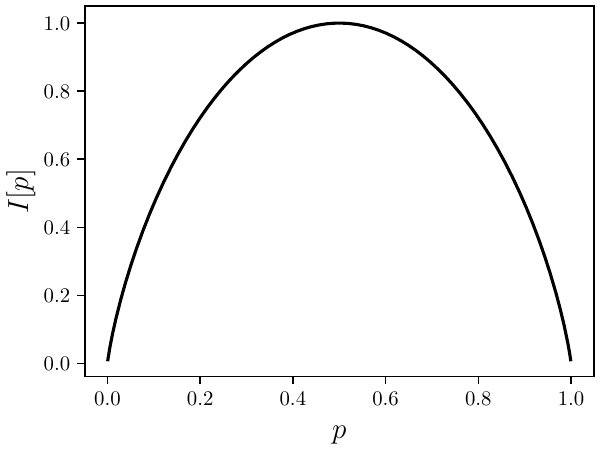}
\end{figure}
Let's look at the general formula. If any of the probabilities are zero, they don't contribute anything to the entropy, because $p \log p$ approaches 0 as $p \to 0$, even though $\log p$ does not. If you have a system in which there is only one state with nonzero probability, then all the $p_{n}$'s will be zero except for one of them. Since they have to add up to one, when $N=2$ the other state will have to have probability 1, and $p \log p = 0$ if $p=0$, but also if $p=1$. So that's why you end up with zero information at either end. This also corresponds to your intuition. What does it mean to be at point $(0,0)$ in the above plot (Fig 1)? It means you've asked a question, and there are two possible answers, but I already know that one of them has probability 0 of being correct, so I know it's the other one. And you can tell me the answer but I won't have learned anything, because I already know what it is, and so the entropy is zero. Symmetrically for the other case. 

{\bf Tarek (TT):} You said you already `know' that the answer is correct, but maybe you would need to think about it for a very very long time. 

{\bf Bill (WB):} Good. One of the things that information theory can do for you, in much the same way as thermodynamics, is it can tell you what's possible and what's not possible, and maybe I'll give some examples. Just as with thermodynamics, however, if I tell you what's possible and not possible, doesn't mean it's realizable, or realizable with limited means. For instance, it might be that I've given you enough data to answer a question, which means that the information content of the data is large enough to resolve whatever question you had in mind, but it might be that to actually do that would require a painfully difficult calculation. And so people talk about information limits, computational limits, storage capacity limits etc. I might somewhere in the process of figuring things out need a lot of space to write things down. There's a whole bunch of ways in which I might be limited, but this is only about information content. And I think it is information content in the colloquial sense, in which we can distinguish the knowability of something from the difficulty of figuring it out. 

{\bf Jakob Metson:} What happens if we don't know these probabilities? It feels like if we know $p$ we know quite a lot about the system already. 

{\bf Bill:}  One cheap answer would be to say the theory is silent. We could go back one step more: what if you didn't know the list of possible answers? I think there's a strong argument to be made that if you can't make a list of the possible answers, then it'll be very difficult to attach a number to how much I learn when I answer it, and so maybe you can't do it. And that's why he [Shannon] chose this setup. I would emphasize that this setup is quite powerful: remember that you can assign probabilities to the best of your knowledge, right? And you can talk about possible answers as being built out of ingredients that can be put together in arbitrary combinations, most of which probably can't happen in the real world, but you don't know that. So you can actually use this in situations that are much broader than the first ones that you think of. There's also a very important point, which is that in this abstraction, to give a list of possible answers is to enumerate them and give pointers. So you might say to yourself wait a second, Shakespeare wrote some number of plays, and it's also true that if I take a section of the newspaper it has some number of news articles. So let me find the right newspaper so that I match Shakespeare's output. So if you tell me which newspaper article it is and you tell me which Shakespeare play it is, how can that be the same amount of information? The plays are much longer! But I didn't ask you what was in the play, I asked you which play it was. And for that there's just a list of some number of them, and all I need to do is give you a pointer. 

{\bf Zhi Zhang:} If all the probabilities sum up to one, why does Shannon entropy consider all probabilities and not just have one less degree of freedom?

{\bf Bill:} That is actually a great question. Which one would you take out?

{\bf Zhi:} It shouldn't matter, right?

{\bf Bill:} Right, and so therefore you shouldn't do it. Since it doesn't matter, it means they're all on equal footing, right? So to pick one of them would be to break a symmetry. And at some point when you're calculating things, you may want to use that symmetry. Let's do a simple example, a baby version of a more interesting calculation. What's the maximum possible entropy? Let
\begin{equation}
I = -\sum_{n=1}^N p_n \log p_n
\end{equation}
If I want to maximize this, there are two ways to do it. One is what you suggested, which is to take the first $N-1$ probabilities as variables and eliminate the last one, because they sum up to one. But there's another way to do it, adding a Lagrange multiplier:
\begin{eqnarray}
\frac{\partial}{\partial p_k} \left[ I - \lambda \left(\sum_{n=1}^N p_n - 1\right) \right] &=& 0 \\
 \frac{\partial}{\partial p_k} \left[ -\sum_{\rm n=1}^N p_{\rm n} \log p_{\rm n} - \lambda \left(\sum_{n=1}^N p_n - 1\right) \right] &=& 0 \\
 -\log p_k - \frac{1}{\ln 2}p_k\frac{1}{p_k} - \lambda &=& 0 \\
\implies \log p_k &=& - \left(\lambda + \frac{1}{\ln 2}\right)
\end{eqnarray}
which is a constant, which means that $p_k$ has to be the same for all $k$, and in order to impose normalization, it has to be $1/N$.
Now if you were to take one $p_k$ out, you would of course get the same answer, but the algebra would be more complicated. So it's useful to try keep all the possibilities on the same footing. 

So what we have seen so far is that if you want to attach a number $I$ to the thing we colloquially call information, you end up with $I$ being the entropy of the distribution:

\paragraph{(A)} $I = S = -k\sum_{n=1}^N p_n \log_2 p_n$. \\

\noindent That's the first great result. 

Let me now do something completely different, which is that I ask you to write down the answer, and I want to know how much space you need to write it out. Let's think about Shakespeare's plays again. If you're taking a course about him, the things you're reading come out of his collected works, and there are standard editions of the collected works: the plays, the sonnets etc. The point is that there is a list; he only wrote so many things. And I need only tell you a number which points to an ordered list of his works. So in order to answer the question ``what play am I reading?'' I don't need to write down the text of the plays, I just need to give you a number between 1 and 250 or so, an unambiguous pointer to the right place in the collection, as long as we both have the same edition. 

Let's assume that we have common knowledge of the language we are going to use to represent the answer. For example, I have eight possible answers. You could say `well, write down a digit from 1 to 8'. How much space do you need? The answer is one digit, but the digit has to have 8 possibilities. If I asked you a question that had 16 possible answers, you could say you only needed one character, because you could represent it in hexadecimal, except nobody likes doing that. And also I changed my alphabet when I changed the question. Not a good idea, it makes everything ambiguous.

To make this question well-posed, we should agree on an alphabet. You can pick whatever you want, but you have to hold it fixed for whatever question you ask. And you don't want to ask how much, you want to ask what is the \textit{minimum} space you need to write the answer, because obviously there's all sorts of ways of writing it down that will waste space. So the alternative, as many of you know, is to agree that our alphabet will be binary. The answer is: it takes three characters. And those are binary digits, so these are also called bits.

{\bf Sheda Ben Nejma:}  The way I am used to think about entropy is our lack of information about a system: we don't know the exact positions of the molecules. To connect with this, should I imagine that I don't know the probabilities when I do this kind of calculation?

{\bf Bill:} No. Let's ask the following question: what is the microscopic state of the gas molecules in this room? So tell me their positions and velocities. How much information will you gain when you see the list of all the positions and velocities? The answer is the entropy of the distribution. So the reason that it gets confusing is that the information we are talking about is the information you \textit{will} gain when you hear the answer, which means that it's also the information you don't have now. So it's the information you don't have that we associate with the entropy usually in statistical mechanics, and that's the connection. But hold on, because we are going to do it more carefully.

{\bf Simone Cicolini (SC):}  So am I correct to think that there is no absolute concept of information, but just information gained relative to my prior knowledge? 

{\bf Bill:} Correct. But I think that corresponds to our intuition, right? When I ask ``how much did I learn?'' it's relative to where I started. I could also project myself backwards to when I knew less, but it's still what I knew at the time. In contexts like this it's misleading, because you can always say `well, imagine you know nothing' which is to say all the answers are equally likely, but that's just an artificial feature of mathematics.

If I have eight possible answers, I can write down the answer using three binary digits, and let's assume that they're all equally likely. Then the information you gain upon hearing the answer is 3 bits. But also, to write the answer, you need three binary digits. So in this trivial example this seems like a coincidence, but it's not. What Shannon proves, and this takes more work, is that if you represent each answer with a codeword that has $l_n$ binary digits, then the average code length is

\paragraph{(B)} $\la l \ra = \sum_{n=1}^N p_n l_n \geq I = -\sum_{n=1}^N p_n \log_2 p_n$ \\

\noindent The fact that it's binary digits is why we chose log base 2 here. It's arbitrary to choose binary for your alphabet, but it was also arbitrary to choose log base 2, so it's a consistent choice. This has many more implications, but I'll tell you about the conceptual one first. If I want to know how much information you gain, I can ask you to write down the answer. If you're efficient in space you use to write down the answer, that space is itself the same measure of how much information you gain. These two very different seeming questions have the same answer. 

The other thing is that it tells you that if you have some complicated probability distribution, say 100 different answers, but the distribution is very funny shaped, so that the entropy still comes out at 3 bits, then there must be some way to encode things so that on average you only need three binary digits, which is surprising because there were so many possibilities! But you all know this, because this is why you're able to watch a movie on your laptop, or even better, on your phone. If I ask you what is the intensity of RGB in every single one of the million pixels on every frame of which there are 25 in a single second, the amount of space it takes to write down all those numbers is enormous, but the probability distribution that they come out of is \textit{extremely} inhomogeneous. So that means that the entropy of that distribution is much less than the logarithm of the number of possible images you can make. 

In particular, let's think about it, even if the images are binary, if I have 10 pixels I can make 1000 images. 20 pixels, one million pictures. An hour is 3600 seconds. And if you have 30 frames per second that's about 100,000 frames. So there are more 20 pixel images than there are frames in a movie. But the movie has images with millions of pixels in them. So something's wrong somewhere, and that is that the probability distribution of these pixels in real movies is incredibly concentrated, so its entropy is much lower. So you can use much less space than you thought would be necessary, and that's the problem of image or video compression, and that's what's driving a reasonable fraction of the technology that we live with today. And Shannon told us what the limit is, the limit is the entropy. By telling you that there's a limit, and that it's smaller than the naive thing, he created the problem of data compression. 

It's worth going back and reading these papers \cite{shannon_48}. It is actually one paper that appears in two parts. There's so many things that happen in it. We often tell the history of sciences as ``on Thursday, this great figure did one thing, and then fifteen years later on Wednesday, that other person did this,'' etc. You know that the selection of people is somewhat biased, and that is problematic in many ways. But I think as working scientists you all also know that this model, that something was discovered on some particular day, before which we didn't know anything, and after which we knew for sure -- that model is also wrong. The history of science is much richer than the one that says ``in this great paper, Higgs wrote down the Higgs model.''  As many of you know, lots of people wrote down the model, and it's an extremely complicated history, and it's much more interesting than that. 

To a remarkable extent, Shannon is one of the cases that conforms to that oversimplified model. Almost everything that was the program of information theory has its roots in one paper, and it's mostly high school mathematics. Subsequent digestions of these ideas have a tendency to be much more complicated. I don't understand why it's often harder to read some of the modern textbooks than it is to read Shannon's original papers, and that's also very unusual. Usually our explanations of things get better with time. 

It is also worth pausing to note that what Shannon did in this argument is very different from our conventional experience in using mathematics to describe the natural world. In most of physics, we have some set of observations (e.g., the motion of planets in the night sky) that can be made quantitative (as Brahe did), and we search for mathematical structures that can explain and unify these data (Kepler, Newton). In contrast, Shannon considered an everyday phenomenon for which we have a colloquial language, and asked if this language itself could be made mathematically precise, without reference to quantitative data. It is remarkable that this worked, and that Shannon's construction has so many consequences.

Proving \textbf{(B)} is substantially harder than \textbf{(A)}, but this is just to give you pointers. If you want to see the proof, go back to the original papers. 

At this point, we've already had a question about the relationship to the idea of entropy in physics. What's interesting about this is that entropy is the answer to the problem of quantifying information, but it's also the answer to how much space it takes to store or represent information. I'll remind you that entropy has its origin somewhere else.

\paragraph{(C)} $dS = \dbar Q/T$ \\

\noindent In thermodynamics, if you keep track of the flow of heat, you have the problem that heat is not a state function. I can talk about the energy content of this piece of chalk, but I can't talk about its heat content. I can say that when I put it in the refrigerator it will cool down, and heat will flow from the chalk into the refrigerator where it will get dissipated somewhere. And if you warm it back up then heat will flow back in, but you're not guaranteed that the amount that flowed out and flowed in balance each other so that there's a well-defined state function called heat content. What was realized, is that if you keep track of heat flow in the right units, the units being the temperature of the bath to which heat from the object is flowing, then you get a state function. So this little slash is to remind you that you can talk about the differential heat, but there is no function called $Q$---if you try to integrate $\dbar Q$, you get nonsense. In particular if you go around a cycle you get something that's not zero which doesn't make any sense for something that's supposed to be a derivative. But there is a state function called $S$, and that's the entropy. 

Then, with the advent of statistical mechanics, which started in the microcanonical ensemble and not in the canonical ensemble, there was this idea of how many microscopic states are accessible. That number was called $W$, and it is one of the great results of statistical mechanics that the logarithm of that number is the entropy.

\paragraph{(D)} $S = k\log W$. \\

\noindent I challenge you to remember how the connection was made in your original statistical mechanics class. This entropy turns out to be the same as \textbf{(C)}. There's a lot of amazing things that happen in physics, this is one of them. What on earth should the number of possible states have to do with heat? Remember, $W$ is not the number of states the system is in, it's the number of states that it \textit{could be in} consistent with the macroscopic constraints that you gave. And that size of the possible phase space apparently has physical content, because it is related to heat flows. There's already something quite surprising here. 

Let me suggest the following thought experiment. Let's measure the positions and velocities of particles in this room. Now take that list of numbers and type it into your computer as a file, and compress it. If you're a Unix user you can just type `compress', otherwise you can use GZip or something similar. Now look and see how long the file is, write down the number. Now warm up the room by ten degrees and go through the exercise again. You will discover that the file is longer. Why is it longer? Well, those compression algorithms are pretty good for this kind of data, so they get close to the Shannon bound on the shortest possible code. So the amount of space those files take up is very close to being the entropy of the distribution out of which those numbers were drawn. But that's the thing we know about from statistical mechanics, which turns out to be the same as the heat flow. This means that if you look at how much longer the computer file is, you can predict how much heat you had to put into the room in order to warm it up by ten degrees. That's the content of Shannon's results about entropy being related to information and code length. 

{\bf Simone:} Suppose I know the position and momentum of every particle in the room, whatever that number may be, how does it not take the same amount of space to store it? 

{\bf Bill:}  ... store them at fixed resolution. So let's agree that you can only measure positions within a micron, as well as velocities and momenta etc. Now do the exercise. So what's going on? When you warm the room up the numbers are all bigger in absolute value. Because you had fixed resolution that means that the numbers are bigger in the absolute sense, so if your resolution is 1 unit of velocity, then the numbers you'll get will be larger, which means you'll need more space to write them down. This actually brings up another problem, which is about going from discrete distributions to continuous distributions. So the reason you're having that difficulty is because it's a bunch of real numbers, but to actually write down a real number would take an infinite amount of space. But the difference between the two could still be finite.

Let me talk a little bit about the more normal case. The case that Shannon started by talking about is you ask a question and you hear the answer. But the more normal case is, unfortunately, if you ask people a question they'll tell you something which is relevant to what you wanted to know, but is not exactly the answer to the question. Does it remove all of your uncertainty? So the original setup was: I have $N$ possible answers, and I go down to one `true' answer, and the $N$ possible answers were associated with a probability distribution $\{p_n\}$. But since I now know that there is one true answer, $p_1=1$, $p_2=p_3=...=p_n=0$. So if I describe the situation before I hear the answer, there's some distribution. Afterwards, when you say you know the answer, no, it's that the distribution of all the possible answers has a very special form: it's 1 and all zeros. The entropy at each extreme is zero, right? $p\log p = 0$ either at $p=0$ or $p=1$. There's only one state, there's no uncertainty so there's no entropy. Here the entropy is what it usually is. And you'll notice that the change in entropy $\Delta S = S_{before} - S_{after} = S_{before}$ is what Shannon identified as being the information gain. 

So without going through the argument, the more general statement is that information gain is entropy loss. This again corresponds to the intuition about the entropy being the information you don't have about the microscopic state of a system. If I could push on the system so that I specify the microscopic state more and more, its entropy would go down, and that reduction in entropy is what you identify as being the information that you gain.

This is important because it is more general. The one where the entropy goes to zero after you hear the answer to your question, that's really weird, and almost never happens. It's useful for an abstract construction, but it never happens in the real world. So this is the general and generally useful statement. 

Let's make an aside. Suppose that I have a probability distribution of some continuous variable. And you say `I don't know what to do about continuous variables, so I am going to make little bins of size $\Delta x$.' And so $p_n$ would be $P({x_n}) \Delta x$. I make these small bins so I don't have to worry about the integral, I just multiply and add. I've turned my continuous problem into a discrete problem, and of course you do this all the time whenever you analyze data numerically. Now I go to compute the entropy of the probability distribution.

If I started at $n=0$ and went all the way to  $n\Delta x = L$ -- so that my bins cover the whole axis -- then what I can do is to send $\Delta x$ to zero because of course that was arbitrary. I really shouldn't have been making bins, there's really an underlying continuous variable, and of course in the process $n$ goes to infinity. Now what happens to the entropy? You all know that if you have a sum of functions evaluated at discrete points and you multiply by the spacing between the points, that becomes the integral. So it becomes 
\begin{equation}
-\sum_n p_n\log p_n = -\sum_n P({x_n}) \Delta x\log \left[P({x_n}) \Delta x\right] \rightarrow
-\int_0^L dx \,P(x) \log_2 \left[P(x) \Delta x\right]
\end{equation}
And that's wonderful, except for this $\Delta x$ term. So if you compute the entropy in the naive way in bins of size $\Delta x$, what you will get is
\begin{equation}
S[P] = -\int_0^L dx \,P(x)\log P(x) - \log \Delta x
\end{equation}
What is this saying? It's saying that if you measure with smaller bins then the entropy is larger, and that's correct. If you're distinguishing more states of the system the entropy goes up. So you can view this in several different ways. You can say, this is a good way of figuring out what dimensionality my data take up. 

{\bf Tasmin Sarkany:} I don't understand something. Haven't you made $\Delta x$ tend to zero, so doesn't it disappear from the expression? 

{\bf Bill:} Right. The point is that the $\Delta x$ which is next to $P(x)$ becomes the $dx$ of the integral, but the $\Delta x$ that is inside the logarithm, you're stuck with it. 

{\bf Tasmin:} But isn't it going towards zero as well? 

{\bf Bill:}  Yes, you let it go towards zero. But when it goes towards zero, you get a finite part, and you get a part that's logarithmically divergent. By the way, this is one of those things where a little bit of simulation goes a long way. Go pick a bunch of Gaussian random numbers, make bins along the axis, turn that into a discrete distribution by making a histogram, compute the entropy, make the bins smaller, and convince yourself that this really happens. It must, because the math is straightforward, but it is interesting to see how well it works. Suppose that we were in two dimensions, so $x$ was a 2-dimensional variable, instead of a 1-dimensional variable. Or even a $D-$dimensional vector. In that case the entropy is
\begin{equation}
S[p] = -\int d^Dx P({\vec x} )\log P({\vec x}) - D\log \Delta x
\end{equation}
That means that if you can isolate this logarithmically diverging part, you can determine the real dimensionality of your data. A nice exercise for instance is to take points, put them on a circle, and then jitter them by a little bit. Now, if you look with $\Delta x$ reasonably large, then they fall on a circle, so they're one-dimensional. Now if you make $\Delta x$ small enough that you can see the jitter, now they're in two dimensions. They don't fill up a 2-dimensional space, but they are a 2-dimensional object. So that means that if you do that exercise and change $\Delta x$, what you'll see is that this thing converges to the right answer pretty quickly, but this piece starts out looking as if $D=1$, and as you make $\Delta x$ smaller and smaller it crosses over to looking like $D=2$. So if I take a bunch of points that are scattered around a circle and I ask you is this 1-dimensional or 2-dimensional, the correct answer is `it depends on how close you look'. And this makes that perfectly precise. Many of you are interested in dimensionality reduction methods, and this is one of the ideas that stands behind that. 

There is however a very important point, which is the entropy is really a functional of the probability distribution. And to be more precise, it's a functional that depends on your resolution. And what that means is, to get back to the problem you were having about the gas molecules in this room, if I don't keep track of my resolution I'm talking nonsense because I have an infinite term. However, if I ask you about $\Delta S = S_{after} - S_{before}$, the limit as $\Delta x$ goes to zero of $\Delta S$ is well-defined. So what that means is when you compute entropy differences, you can go to the continuum limit and there's no problem. Continuous variables are very annoying when you want to talk about entropy, but as long as you only talk about entropy \textit{differences}, there's no problem.

Now let me remind you, in classical statistical mechanics, phase space is continuous. Remember it is positions and momenta. That means that strictly speaking, \textit{the} entropy is not defined until someone gives you a natural $\Delta x \Delta p$. You will remember that around 1900 we got a natural $\Delta x \Delta p$, it's called Planck's constant. But classical statistical mechanics was very well developed before Planck, so what's going on? The answer is that the absolute entropy never appears in your discussion. You measure entropy changes, because they are associated with heat flows. And those are perfectly well defined in the continuum limit. And that's why you can have a subject called classical statistical mechanics in retrospect. Obviously that's not how they got there. 

{\bf Amin Safaeesirat:} Is there any proof on the equality between \textbf{(B)} and \textbf{(C)}, and \textbf{(C)} and \textbf{(D)}? 

{\bf Bill:} Well, \textbf{(C)} and \textbf{(D)} is the relationship between statistical mechanics and thermodynamics. If you believed that what you were doing in your statistical mechanics class is related to what you were doing in your thermodynamics class, you had better believe that these are the same $S$. Now, there's these two different physical pictures. That's the content of statistical mechanics. The thermodynamic quantities that you compute as expectation values in stat mech are the thermo quantities that you measure. This is a kind of hypothesis. There's a proof that \textbf{(B)} is the same function as \textbf{(D)}, it's just math. In one case it's a list of possible states or signals of messages or answers to the question, and in another it's a list of microscopic states of a macroscopic system. 

{\bf Natalie Blot:} So if you have an equivalence between information and entropy, then you have an equivalence between information and energy ... 

{\bf Bill:} Careful ... 

{\bf Natalie:}  ... and there's an equivalence between energy and mass, so when you change a hard drive's contents do you also change its mass? 

{\bf Bill:} So I take a polymer. I could take the gas in a room or the gas in a balloon, but let's have a little more fun, it's a biophysics course. So let's take a single strand of DNA or a double strand of DNA, and I pick up both ends in optical tweezers, and I don't stretch them out too far, and I pull. As I pull, I restrict the conformation of the polymer. The number of ways the polymer can get from here to there given its length goes down, so the entropy of the polymer goes down by an amount that you can at least try calculating using the random walk polymer -- you know it's self-avoiding, you know it's more complicated because it can twist, but if you did the random walk polymer that wouldn't be terrible. Now because you haven't actually stretched any bonds, the energy of the system has not changed. The \textit{free energy} has changed, because the entropy has changed. And since the free energy has changed, you needed to do work in order to pull the two beads apart, but that's free energy not energy. In order to induce an entropy change, I might need to do work, in which case energy changes hands, but strictly speaking they are independent objects. In particular there's a conversion factor which is the temperature. The relatedness of entropy and energy is mediated by the temperature of the surroundings. There's a whole field of thinking about the energetic costs of manipulating information and the short answer is -- and this goes back to Maxwell's demon -- as long as you don't destroy any information, as long as every transformation you make can be inverted, you don't need to spend any energy.

{\bf Natalie:} Spend energy in the sense of do work. 

{\bf Bill:} Right, or dissipate heat. The place where you end up having to dissipate energy is when you erase something, so when you lose information irreversibly. But you don't need to think about that in order to understand this, and I don't think it'll be important for anything that I want to talk about. 

This is to remind you that entropy differences are perfectly well-defined. That means the information you gain about a continuous variable by measuring it is perfectly well-defined, despite the fact that if I ask you for the entropy of the continuous variable you'll be in trouble. So let me do one calculation, and then point out some consequences, and then we'll stop. Let's pick up this idea of using continuous variables. Let me very carefully consider the following situation. $y$ is what you can measure, and $x$ is what you care about. $\xi$ is noise, and to be precise, let's assume that it's Gaussian with variance $\sigma^2$.
\begin{equation}
	\label{eq:fun_noise}
y = x + \xi
\end{equation}
So what you get to observe is the thing you really care about but, unfortunately, with a Gaussian random number added to it. So the question is: by observing $y$, how much information do you gain about $x$? We know how to do this, it's
\begin{equation}
I_{y \to x} \equiv S[P_X(x)] - S[P(x|y)] .
\end{equation}
And then you could say, that's what I would get if I observe a particular $y$. But maybe I want to know what I would get if I did this many times and averaged over $y$. Then, I want to think about the average of this quantity by integrating it over all possible $y$s weighed by their probability of occurrence,
\begin{align}
\la I_{y \to x} \ra_y &= \int dy\, P_Y(y) \left( S[P_X(x)] - S[P(x|y)] \right) \\
&= \int dy\, P_Y(y) \left[- \int dx\, P_X(x)\log P_X(x) + \int dx\,P(x|y)\log P(x|y) \right] .
\end{align}

Now, you'll notice something interesting. There's a $P_Y$ factored out of the brackets, and the leftmost term in the brackets has a $P_X$, but no $P_Y$. From the point of view of the integral over $y$, that term is a number. If I wanted to, I could `cheat' and condition on $y$. I could do the integrals in the other order, and that would lead to the same result.
\begin{align}
\la I_{y \to x} \ra_y &= \int dy\, P_Y(y) \int dx\, P(x|y)\log \left[\frac{P(x|y)}{P_X(x)} \right] \\
&= \int dy \int dx\, P (x|y)P_Y(y)\log \left[\frac{P(x|y)P_Y(y)}{P_X(x)P_Y(y)} \right]\\
&= \int dy \int dx\, P (x,y)\log \left[\frac{P(x,y)}{P_X(x)P_Y(y)}\right] \\
&\equiv I[X;Y]
\end{align}
What we've shown is that if I ask you about the average information that $y$ gives about $x$, that's a quantity that turns out to be symmetric in $x$ and $y$, and it has a name: it's called the mutual information. It's mutual because it's symmetric, and it's a kind of generalized correlation. 

What does ``generalized correlation'' mean? If you see data that looks like this: 
\begin{figure}[H]
\centering
\includegraphics[width=3in]{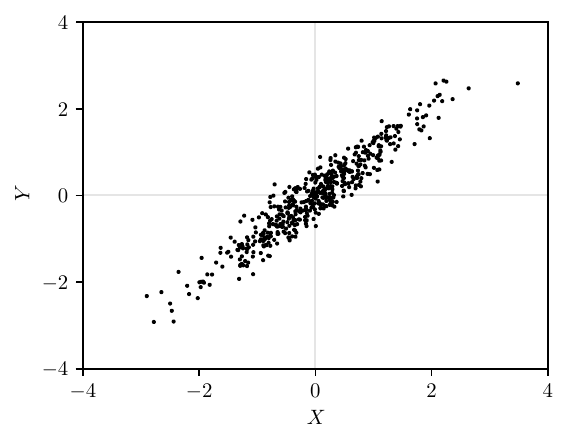}
\end{figure}
\noindent you can measure the correlation coefficient $C(x,y)$, and that will tell you that $x$ and $y$ are related to each other. In the above, $C(x, y) \neq 0$, and $I[x;y] \neq 0$. But if you see this: 
\begin{figure}[H]
\centering
\includegraphics[width=3in]{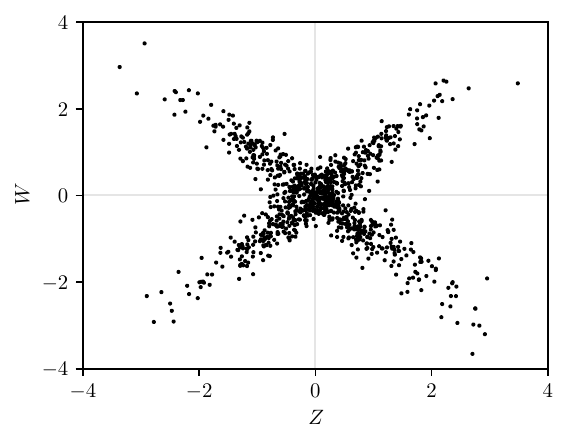}
\end{figure}
\noindent the correlation coefficient is zero. But you would not want to conclude that $z$ and $w$ are unrelated! In fact, they are almost as related as $x$ and $y$, there's just a twofold ambiguity: you don't know whether you are on one line or the other. So in fact, $I[z;w] \sim I[x;y] - 1$, just one bit smaller, because you have just one more yes/no question you need to answer, which is whether you fall on this or that line. 

One of the things mutual information does for you is give you a measure of relatedness among variables that generalizes the usual algebraic notion of correlation. In fact, since information theory started by talking about discrete variables, you don't even need a metric. Mutual information is independent of units of $x$ and $y$. That's true of the correlation coefficient, but not, for example, about the covariance. That's interesting, it tells you how $x$ and $y$ are related to each other without needing to know the units in which $x$ and $y$ are measured. That's good. But it's stronger than that, because it's also invariant under any invertible transformation. So if you decide that instead of $x$, you want to measure $e^x$, it's fine, it doesn't change the mutual information with $y$. That's quite deep, and it's not true of any metric or algebraic measure of correlation. 

You can be more formal about this, that if $x$ and $y$ are themselves multidimensional you can think about doing all sorts of complicated nonlinear transformations of the space, but as long as it's invertible then everything's okay. It's because it's about information! And as long as everything's invertible, you don't lose any information. I think this is actually quite interesting, that not only do you start with something that was discrete, you try to go to the continuum limit and you make a mess because of the $\log$, but then if I ask you about mutual information everything's fine again because that's an entropy \textit{difference}. In fact you automatically grow not only a finite continuum limit, but you get complete invariance to any separate invertible transformations of the two variables. That happened for free, you didn't insist on that being a property of your measure. 

Now let's do one real calculation. Let me consider a probability distribution $Q(y)$ which is Gaussian, and let me compute the entropy of this thing not worrying about the diverging $\log$.
\begin{align}
S[q(y)] &= - \int dy\, Q(y)\log Q(y) \\
&= - \int {{dy}\over {\sqrt{2\pi\la y^2 \ra}}} e^{-\frac{y^2}{2\la y^2 \ra}}\left[-\frac{1}{2}\log \left(2\pi \la y^2 \ra\right) -  \frac{y^2}{2\la y^2 \ra}\log e \right] \\
&= \frac{1}{2}\big [\log \left(2\pi \la y^2 \ra\right) + \log e \big ] \\
&= \frac{1}{2}\log \left(2\pi e\la y^2 \ra\right)
\end{align}
Why did I do this? We said that $y = x+ \xi$, and $\xi \sim \mc{N}(0, \sigma^2)$ (see Eq. \eqref{eq:fun_noise}). One way I could write the mutual information is as a difference between entropies. We did it before as
\begin{equation}
I[x;y] = S[P_X(x)] - S[P(x|y)]
\end{equation}
but it's symmetric, so I could also write it as
\begin{equation}
I[x;y] = S[P_Y(y)] - S[P(y|x)]
\end{equation}
From this picture, $P(y|x)$ is Gaussian. So the rightmost term in the above equation is the one we just calculated, which is why we calculated it. What about the left term? Well, in general you can't say anything, because the distribution of $y$ depends on the distribution of $x$. But I can tell you something. I can tell you that
\begin{equation}
\la y^2 \ra = \la x^2 \ra + \sigma^2
\end{equation}
And I can now ask you for the following quantity. What is the {\em maximum} over all distributions of $y$ of the entropy of that distribution, with the constraint that the variance is known? The answer is that the probability distribution that has the largest possible entropy, given you know its variance, is a Gaussian. 

You will remember from your statistical physics course, one of the ways of thinking about the Boltzmann distribution is it is the probability distribution that has the largest possible entropy given that you know the average value of the energy. This is the same kind of idea, you find the probability distribution that has the maximum possible entropy given the average of $y$ and the average of $y^2$, and you'll find that it's a Gaussian. If I had a little more time I would do the calculation for you, but it's not hard. What that means is,
\begin{align}
S[P_Y(y)] &\leq  \frac{1}{2}\log \left[2\pi e \left(\la x^2 \ra + \sigma^2 \right)\right]
\end{align}
Now we are almost done. What we've shown is that
\begin{align}
I[x;y] &\leq \frac{1}{2}\log \left[2\pi e \left(\la x^2 \ra + \sigma^2 \right)\right] - \frac{1}{2}\log 2\pi e \sigma^2 \\
&= \frac{1}{2}\log \left(1 + \frac{\la x^2 \ra}{\sigma^2} \right)
\end{align}

So, if you're measuring a signal that's embedded in a background of noise that has variance $\sigma$, and you measure the variance of the signal itself, and you take the ratio, that's the signal-to-noise ratio. Nobody told you what units to measure the signal or the noise in, what matters is the relative magnitude. And since they're fluctuating, every time you look $x$ has a different value, to measure their scale you compute their variance. You take the ratio of their variances that gives you a measure of how big the signal is in units of the noise. Let's do a simple exercise. Suppose the typical scale looks like this: 
\begin{figure}[H]
\centering
\includegraphics[width=3in]{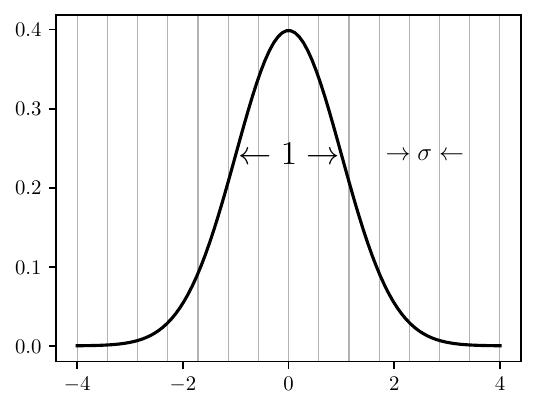}
\end{figure}
\noindent We'll use units in which that's about 1. We're going to assume that $\sigma$ has some very small width. What does that mean? Roughly speaking, of course you should do a real calculation, think about noise, noise distributions have tails, so you can get confused about things even if they're 1 sigma apart, but roughly speaking, the noise sets a scale, and if the variance of the signal is 1, then the number of bins here gives $1/\sigma$. So if I ask you how much information do you get, the answer is that by making measurements that have noise level $\sigma$, you can distinguish roughly $1/\sigma$ different possibilities which have roughly the same probability because they're in the bulk of the distribution. That means you'd guess that the amount of information you should get should be $\log (1/\sigma )$. Well, if $\la x^2 \ra$ is 1, and $\sigma$ is a small number, and if we ignore the 1 inside the $\log$, then this is what the mutual information is telling us. It is saying rigorously the thing that you intuitively feel, which is that if the signals you are looking for are of scale 1, and the noise is of scale $\sigma$, then you can distinguish $1/\sigma$ different possibilities. 

That means by the way, that in order to get 10 bits, $1/\sigma$ has to be $1000$, which means you have to make measurements with an accuracy of $1/1000$ of the scale of the thing you're looking for. 10 bits is a lot!

\section{Afterword}

As it turned out, this tutorial was  an introduction to the basic quantities of information theory and their relations to things that are more familiar in statistical mechanics and thermodynamics.  There really wasn't time for much more.
These  ideas have many uses in thinking about the physics of life.  Examples include:
\begin{itemize}
\item Making direct estimates of the amount of information available in sequences of action potentials, in the concentrations of signaling molecules, etc..  Such estimates quantify our intuitions about signals and noise, and allow us to dissect which features of the signal are most relevant, testing different models for the underlying code.
\item Connecting the information available with the information needed in order to accomplish various functions essential for the organism.  While evolution does not select for collecting or transmitting bits, reaching criterion levels of performance (and, ultimately, fitness) always requires a minimum amount of information.
\item Finding limits to the information that can be represented with limited physical resources---numbers of molecules, numbers of action potentials, ... .
\item Testing whether real biological systems come close to these limits.
\item If the physical limits to information transmission are relevant, developing theories that define what is needed for this near--optimal performance.
\end{itemize}
The last step is the most ambitious, but getting to this stage rests on all the previous steps.

\paragraph{Funding information.}
WB was supported in part by the US National Science Foundation through the Center for the Physics of Biological Function (PHY--1734030), and by Fellowships from the Simons Foundation and the John Simon Guggenheim Memorial Foundation. TT was supported by the Agence Nationale de la Recherche (ANR) DISTANT grant.

\end{document}